\documentclass[a4paper,11pt]{article}
\pdfoutput=1 % if your are submitting a pdflatex (i.e. if you have
\usepackage{jinstpub} % for details on the use of the package, please
\usepackage{lineno}
\usepackage{subfig}
\usepackage{multirow}
\title{\boldmath Study of cosmogenic activation in $^{76}$Ge enriched germanium detectors during fabrication and transportation above ground}

\author[a]{Qiyuan Nie,}
\author[a]{Zhi Zeng,}
\author[a,1]{Hao Ma,\note{Corresponding author.}}
\author[a]{Litao Yang,}
\author[a]{Qian Yue,}
\author[a,b]{Jianping Cheng}
\affiliation[a]{Key Laboratory of Particle and Radiation Imaging (Ministry of Education) and \\Department of Engineering Physics, Tsinghua University, Beijing 100084, China}
\affiliation[b]{Beijing Normal University, Beijing 100875, China}

\emailAdd{mahao@tsinghua.edu.cn}

\abstract{Rare event search experiments using germanium detectors are operated in underground laboratories to minimize the background induced by cosmic rays. However, the cosmogenic activation in germanium crystals on the ground during fabrication and transportation generates long half-life radionuclides and contributes a considerable background. We simulated the production rates of cosmogenic radionuclides in germanium and calculated the specific activities of cosmogenic radionuclides according to the scheduled fabrication and transportation processes of $^{76}$Ge enriched germanium detectors. The impact of cosmogenic background in germanium crystals for the next generation CDEX experiment was assessed with the scheduled exposure history above ground. }

\keywords{Detector modelling and simulations I, Double-beta decay detectors}

\begin{document}
\maketitle
\flushbottom

\section{Introduction}
\label{sec:intro}

Rare event search experiments, such as the direct detection of dark matter and the search for neutrinoless double beta decay ($0\upsilon \beta \beta$), are operated in underground laboratories with passive and active shields \cite{b1,b2,b3,b4,b5}. Ultrapure materials should be carefully selected for detector fabrication to minimize their intrinsic background. In addition to the contamination from primordial nuclides, cosmogenic radionuclides in the detector’s materials produced via cosmogenic activation can contribute a non-negligible background \cite{b6,b7}.

The China Dark Matter Experiment (CDEX), to search for dark matter and $0\upsilon \beta \beta$ decay of $^{76}$Ge, operates HPGe detectors at the China Jinping Underground Laboratory (CJPL) \cite{b8,b9,b10,b11,b12}. CDEX is proceeding to the next phase of the experiment with 300 kg high-purity germanium (HPGe) detectors. Many efforts have been invested to reduce the background from various radionuclides \cite{b7,b13}. However, the cosmogenic activation of germanium above ground will generate long-lived radioisotopes like $^3$H and $^{68}$Ge. Previous studies have reported a crucial background contribution introduced by these cosmogenic radionuclides in germanium \cite{b14,b15,b16,b17}. It is necessary to assess the background contribution from cosmogenic radionuclides in germanium for the CDEX experiment, which depends on the exposure time of germanium above ground, fabrication locations and shipping routes.

In this study, we present the calculation of the production rates of cosmogenic radionuclides in germanium. Furthermore, we simulate the background spectra and assess the background contribution from cosmogenic radionuclides in germanium crystals of HPGe detectors used in the future CDEX experiment according to scheduled fabrication and transportation processes above ground.

\section{Cosmogenic activation in germanium}
\label{sec.II}
\subsection{Calculation method}

The production rate $R_i$ of a cosmogenic radionuclide $i$ can be expressed as shown in eq.~\eqref{eq1}.
\begin{equation}
\label{eq1}
R_i = \sum_{j}N_j\int\Phi_k(E)\sigma_{ijk}(E)dE,
\end{equation}
where $N_j$ denotes the number of target germanium isotope $j$. $\Phi_k$ is the flux of cosmic-ray particle $k$. In this study, we mainly considered cosmic-ray neutrons, protons, gammas, and muons. $\sigma_{ijk}$ is the cross-section of radionuclide $i$ generated by the interactions between cosmic-ray particle $k$ and target particle $j$. The atomic abundances of our enriched germanium are 86.6\% of $^{76}$Ge, 13.1\% of $^{74}$Ge, 0.2\% of $^{73}$Ge, and 0.1\% of $^{72}$Ge \cite{b14}.

\begin{figure}[htbp]
\centering
\includegraphics[width=.8\hsize]{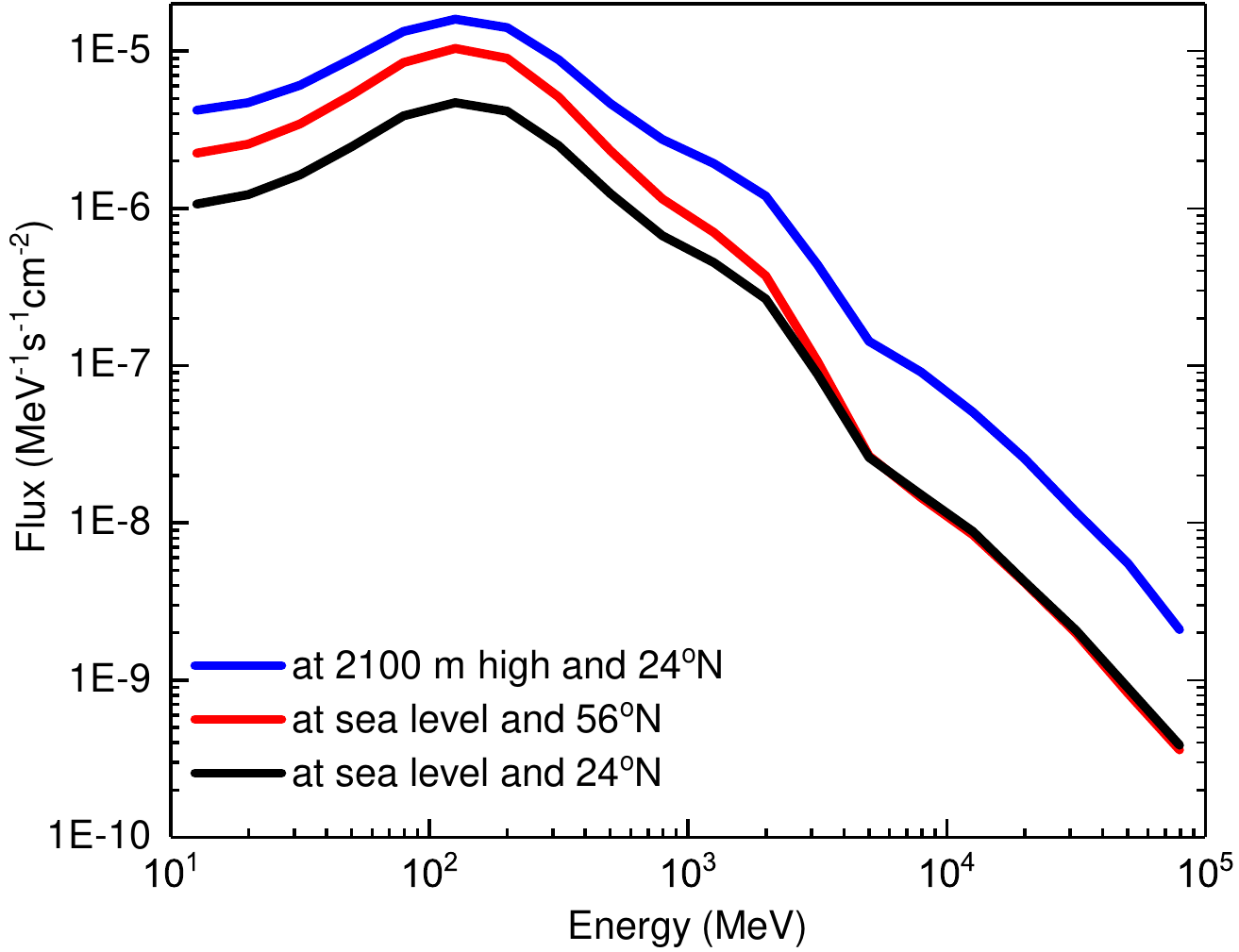}
\caption{Spectra of cosmic-ray neutrons at different locations from the CRY-1.7 library.}
\label{fig1}
\end{figure}

In this study, Geant4 version 10.6.01 with the physics list Shielding \cite{b18,b19,b20} was applied to simulate the interactions between cosmic rays and germanium isotopes. The CRY-1.7 library \cite{b21} can provide fluxes and spectra of cosmic rays. For example, spectra of cosmic-ray neutrons at different locations are shown in figure~\ref{fig1}. Taking the information of cosmic rays provided by the CRY-1.7 library as the inputs of the Geant4 simulation program, we can calculate the production rates of cosmogenic radionuclides in germanium.

The CRY-1.7 library only provides information of cosmic rays at three altitudes of 0 m, 2100 m, and 11300 m. All locations involved in this work have elevations below 2100 m, therefore we set the altitude to 0 m in the simulation and corrected the production rates based on the cosmic-ray fluxes at different altitudes. The cosmic-ray flux at a certain altitude can be calculated by eq.~\eqref{eq2} \cite{b22}.
\begin{equation}
\label{eq2}
\Phi_k(H) = \Phi_k(0)e^{\frac{p(H)-p(0)}{\lambda_k}},
\end{equation}
where $\Phi_k(H)$ and $\Phi_k(0)$ are the fluxes of cosmic-ray particle $k$ at altitude $H$ and sea level. $p(H)$ is in g/cm$^{2}$ and represents the atmospheric pressure at altitude $H$. $\lambda_k$ indicates the typical absorption length of cosmic-ray particle $k$ at low altitudes \cite{b22}. As the normalized spectra of cosmic-ray neutrons are almost consistent at different altitudes within 20 km, the production rates of cosmogenic radionuclides induced by cosmic-ray neutrons are proportional to the intensity of neutron flux \cite{b23}. It is generally considered that neutron dominates the production rates of the cosmogenic radionuclides \cite{b13,b14}, so we ignored the variation of the normalized spectra of other cosmic rays with altitude. If $R_k(0)$ is the production rate of a cosmogenic radionuclide activated by cosmic-ray particle $k$ at sea level and $R(H)$ represents the total production rate at altitude $H$, they can be organized as shown in eq.~\eqref{eq3}.
\begin{equation}
\label{eq3}
R(H) = \sum_{k}[R_k(0)e^{\frac{p(H)-p(0)}{\lambda_k}}],
\end{equation}

\subsection{Production rates of cosmogenic radionuclides}

The fabrication and transportation processes of HPGe detectors are shown in table~\ref{tab1}.
The fabrication of HPGe detectors follows a series of procedure. 
The enrichment in $^{76}$Ge for CDEX detectors is performed in Electrochemical Plant (ECP) in Zelenogorsk, Russia. Germanium isotopes are separated in the form of GeF$_4$ gas. The $^{76}$Ge enriched GeF$_4$ gas is then extracted from the centrifuge and converted into GeO$_2$ powder. This GeO$_2$ powder is transported by ground to Kunming, China, where the reduction of GeO$_2$ to metallic germanium is performed. The produced metal ingots then undergo preliminary zone-refining and etching.
For further purification and crystal growth, these germanium ingots are sent to Oak Ridge, USA. The enriched germanium is further zone-refined and grown into HPGe crystals with specific dimensions. 
In the final step, the germanium crystal slices are fabricated to HPGe detectors in Strasbourg, France.

\begin{table}[htbp]
\centering
\caption{Fabrication and transportation processes of germanium materials and detectors.}
\label{tab1}
\smallskip
\resizebox{\textwidth}{!}{
\begin{tabular}{|l|c|c|c|c|c|}
\hline
    Event & Site or Transportation & Latitude & Altitude (m) & Duration (d) & Shielding condition \\

\hline
    $^{76}$Ge enrichment  & Zelenogorsk & 56$^\circ$N & 300 & 2.2 & none \\ 
    GeO$_2$ powder storage & Zelenogorsk & 56$^\circ$N & 300 & 122 & underground storage \\ 
    GeO$_2$ shipped to Kunming & Zelenogorsk$\rightarrow$Kunming & 26$^\circ$N$\thicksim$56$^\circ$N & 0$\thicksim$2000 & 30 & transportation shield \\ 
    GeO$_2$ converted to Ge metal & Kunming & 26$^\circ$N   & 1500 & 6.5 & none \\ 
    GeO$_2$ shipped to Oak Ridge & Kunming$\rightarrow$Oak Ridge & 26$^\circ$N$\thicksim$52$^\circ$N & 0$\thicksim$2000 & 60 & transportation shield \\ 
    Crystal growth & Oak Ridge & 36$^\circ$N & 300 & 15 & none \\ 
    Crystal characterization & Oak Ridge & 36$^\circ$N & 300 & 15 & temporary underground storage \\ 
    Crystal shipped to Strasbourg & Oak Ridge$\rightarrow$Strasbourg & 36$^\circ$N$\thicksim$48$^\circ$N & 0$\thicksim$400 & 40 & transportation shield \\ 
    HPGe detector production & Strasbourg & 48$^\circ$N & 150 & 30 & temporary underground storage \\ 
    HPGe detector shipped to CJPL & Strasbourg$\rightarrow$CJPL & 28$^\circ$N$\thicksim$52$^\circ$N & 0$\thicksim$2000 & 25 & transportation shield \\ 
\hline
\end{tabular}
}  
\end{table}

During above procedure, temporary storage of germanium in shallow underground locations during non-working hours can reduce the daily exposure time from 24 hours to about 8 hours. These underground locations are within 30 km from the factories, with overburdens exceeding 50 meters of water equivalent. A detailed introduction of these underground storage locations can be found in ref. \cite{b17}.

\begin{figure}[htbp]
\centering
\includegraphics[width=.8\hsize]{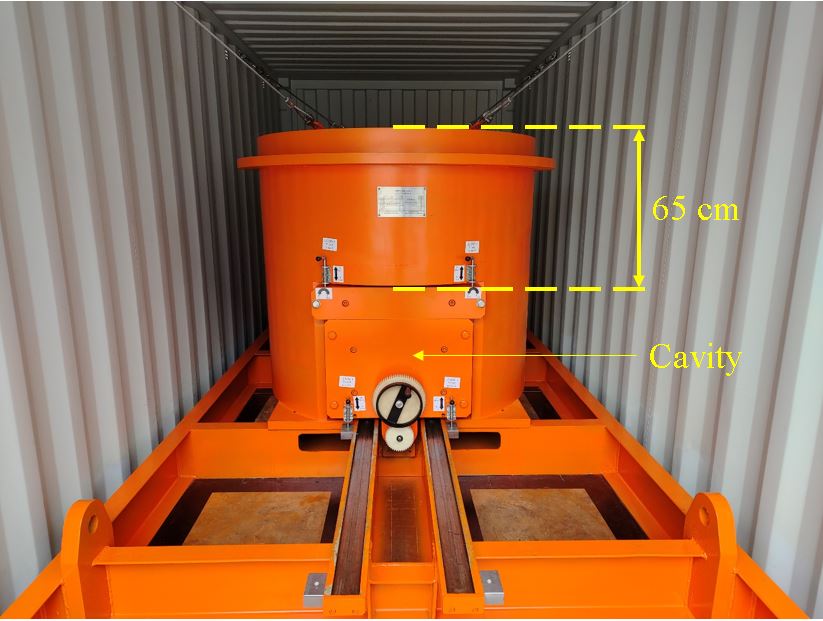}
\caption{The transportation shield inside a shipping container.}
\label{fig2}
\end{figure}

The transportation of germanium is time-consuming and spans continents. It is necessary to apply a shield during the transportation processes to reduce the production of cosmogenic radionuclides \cite{b24,b25}. Steel is an optimal material for transportation shielding, which can shield cosmic-ray neutrons well with a low production rate of secondary neutrons \cite{b20,b26}. A transportation shield made of low-carbon steel was manufactured with a height of 118 cm and a diameter of 139 cm as shown in figure~\ref{fig2}. The shield has an 187 L cylindrical cavity to contain about 35 kg GeO$_2$ powder or 21 typical HPGe crystals. The cavity is covered with 65 cm thick steel above and 36 cm thick steel around it. The transportation shield has a total weight of 14 t to balance the transportation requirements of germanium and the load restrictions in Europe.

The simulated results of production rates of cosmogenic radionuclides in germanium during the fabrication and transportation processes are listed in table~\ref{tab2} and table~\ref{tab3}, with the latitudes ranging from 26$^\circ$N to 56$^\circ$N and the altitudes varying form sea level to 2000 m. Due to the high altitude, the production rates of cosmogenic radionuclides in Kunming are the highest among all fabrication sites as shown in table~\ref{tab2}.

The transportation shield can effectively suppress the production of cosmogenic radionuclides as shown in table~\ref{tab3}. Under the protection of the transportation shield, the production rates of $^{68}$Ge and $^{55}$Fe are reduced by a factor of 8 and 14 at 24$^\circ$N, respectively. As cosmic rays vary with the latitude, the reduction factor of the $^{68}$Ge production is simulated to about 10 at 56$^\circ$N. 
For different cosmic-ray particles, gamma rays are almost completely shielded. The production rates of $^{68}$Ge induced by neutrons and protons are reduce by a factor of 11 and 4 at 24$^\circ$N. Cosmic-ray muons will generate abundant secondary neutrons when passing through the shield, which further increases the production rates of cosmogenic radionuclides. In general, the shield can effectively mitigate the cosmogenic activation in germanium by a factor of 10.

\begin{table}[htbp]
\centering
\caption{Production rates of cosmogenic radionuclides in germanium at different fabrication sites.}
\label{tab2}
\smallskip
\begin{tabular}{|l|c|cccc|}
\hline
    \multirow{2}{*}{Radionuclide} &
    \multirow{2}{*}{$T_{1/2}$} &
    \multicolumn{4}{c|}{Production rate (kg$^{-1}$d$^{-1}$)} \\
    \cline{3-6}
    & & Zelenogorsk & Kunming & Oak Ridge &    Strasbourg\\
\hline
    $^{68}$Ge &270.9 d & 17.17  & 22.47  & 10.69  & 13.45  \\ 
    $^{60}$Co &5.3 yr & 1.51  & 2.49  & 1.11  & 1.23  \\    
    $^{3}$H   &12.3 yr & 39.16  & 70.74  & 29.25  & 31.89  \\
    $^{65}$Zn &243.9 d & 15.17  & 21.97  & 10.08  & 12.33  \\
    $^{55}$Fe &2.7 yr & 5.13  & 7.86  & 3.45  & 4.01  \\ 
    $^{63}$Ni &101.2 yr & 2.52  & 4.65  & 1.95  & 2.19  \\ 
    $^{57}$Co &271.7 d & 2.89  & 5.07  & 2.14  & 2.30  \\ 
    $^{54}$Mn &312.2 d & 0.89  & 1.59  & 0.65  & 0.76  \\ 
    $^{49}$V &330.0 d & 1.09  & 2.23  & 0.93  & 0.91 \\ 
\hline
\end{tabular}
\end{table}

\begin{table}[htbp]
\centering
    \caption{Production rates of cosmogenic radionuclides in germanium generated by different components of cosmic rays with the transportation shield. The simulated results without the shield are listed in parentheses for comparison.}
    \label{tab3}
    \smallskip
    \begin{tabular}{|l|c|ccccc|}
    \hline
    \multirow{2}{*}{Radionuclide}  & \multirow{2}{*}{Latitude} &\multicolumn{5}{c|}{Production rate (kg$^{-1}$d$^{-1}$)} \\
    \cline{3-7}
    & &Neutron & Proton & Muon & Gamma-ray & Total \\ 
    \hline
        $^{68}$Ge & 24$^\circ$N & 0.51(5.62) & 0.18(0.78) & 0.12(0.04) & 0(0.07) & 0.81(6.51) \\ 
        ~ & 56$^\circ$N & 0.99(11.98) & 0.20(1.28) & 0.15(0.05) & 0(0.09) & 1.34(13.40) \\ 
        $^{60}$Co & 24$^\circ$N & 0..02(0.48) & 0.02(0.19) & 0.02(0.01) & 0(0.01) & 0.06(0.69) \\ 
        $^{3}$H & 24$^\circ$N & 0.65(12.01) & 0.56(5.31) & 0.66(0.31) & 0(0.09) & 1.87(17.72) \\ 
        $^{65}$Zn & 24$^\circ$N & 0.35(4.95) & 0.15(1.07) & 0.17(0.07) & 0(0.08) & 0.67(6.17) \\ 
        $^{63}$Ni & 24$^\circ$N & 0.11(1.64) & 0.06(0.46) & 0.08(0.03) & 0(0.03) & 0.25(2.16) \\ 
        $^{57}$Co & 24$^\circ$N & 0.03(0.80) & 0.03(0.42) & 0.03(0.02) & 0(0.01) & 0.09(1.25) \\ 
        $^{55}$Fe & 24$^\circ$N & 0.04(0.89) & 0.04(0.47) & 0.02(0.01) & 0(0.01) & 0.01(1.38) \\ 
        $^{54}$Mn & 24$^\circ$N & 0.02(0.25) & 0.01(0.12) & 0.01(0.004) & 0(0.002) & 0.03(0.41) \\ 
        $^{49}$V & 24$^\circ$N & 0.01(0.35) & 0.02(0.23) & 0.01(0.004) & 0(0.001) & 0.04(0.59) \\ 
    \hline
    \end{tabular}
\end{table}

\subsection{Specific activities of cosmogenic radionuclides} 

With the simulated production rates, we calculated the specific activities of cosmogenic radionuclides in each stage of the fabrication and transportation processes. The specific activity $A_{(i,n)}$ of a cosmogenic radionuclide $i$ at the stage $n$ can be calculated in eq.~\eqref{eq4}:
\begin{equation}
\label{eq4}
A_{(i,n)} = \sum_{\Delta L}[R_i(L,H)(1-e^{-\lambda_i t_L})]+A_{(i,n-1)}e^{-\lambda_i t_L},
\end{equation}
where $R_i(L,H)$ is the production rate of cosmogenic radionuclides $i$ at latitude $L$ and altitude $H$. In the fabrication stage, $H$ is the altitude of the fabrication site. During the transportation stage, $H$ is taken as the altitude of the city or road on the transportation route. Shielding conditions at different stages have been considered. $\lambda_i$ is the decay constant of cosmogenic radionuclides $i$. $t_L$ is the residence time of germanium materials around latitude $L$ in the current stage.

\begin{figure}[htbp]
\subfloat{
\label{fig3a}
\includegraphics[width=.5\hsize]{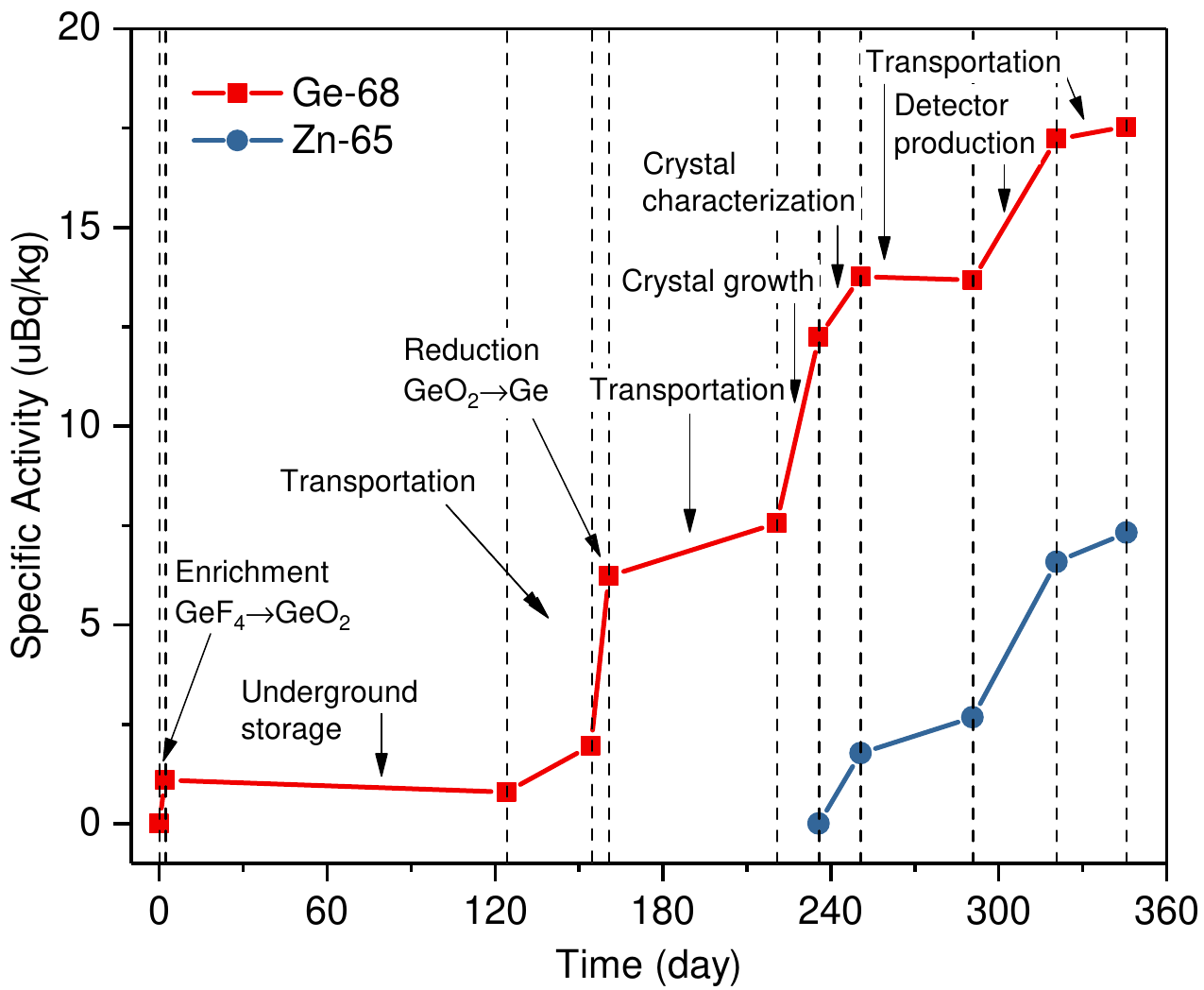}
}
\subfloat{
\label{fig3b}
\includegraphics[width=.5\hsize]{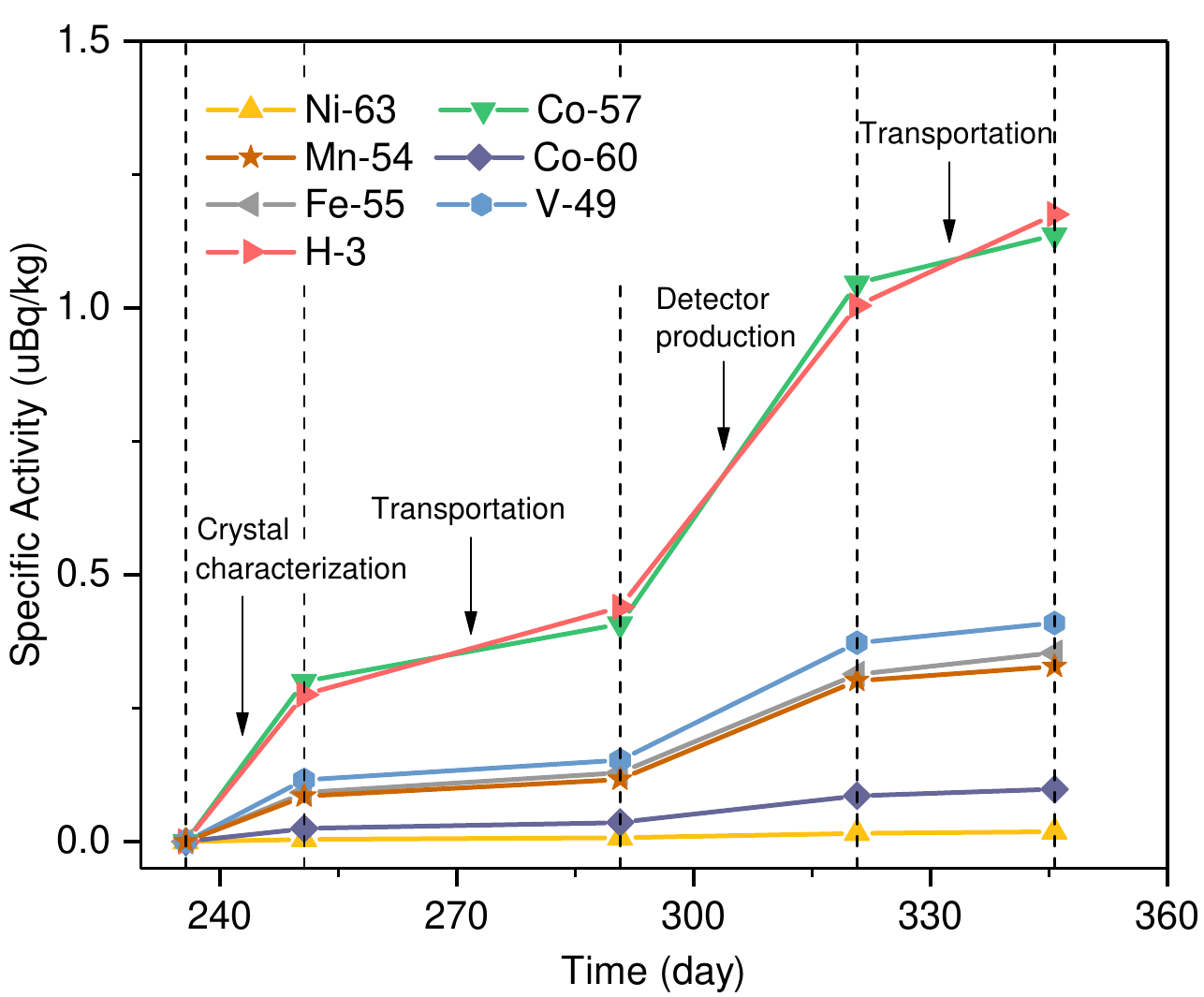}
}
\caption{The specific activities of cosmogenic radionuclides in germanium at different stages of the fabrication and transportation processes.}
\label{fig3}
\end{figure}

The specific activities of cosmogenic radionuclides in germanium during the fabrication and transportation processes are calculated as shown in figure~\ref{fig3}. 
The first stage is the conversion process from GeF$_4$ gas to GeO$_2$ powder.
Cosmogenic radionuclides starts to accumulate right after GeF$_4$ gas leaves the centrifuge \cite{b14,b17}. $^{68}$Ge continuously accumulates and decays during subsequent storage, transportation, and fabrication stages. 
Due to the high altitude in Kunming, $^{68}$Ge accumulates rapidly when GeO$_2$ power is reduced to germanium metal. The specific activity of $^{68}$Ge increases by 4.27 $\mu$Bq/kg within 6.5 days.
Nuclides such as $^{60}$Co and $^{3}$H are almost removed during the crystal growth stage and restart accumulation in subsequent stages. 
During the crystal characterization and detector production stages, the storage of germanium in underground locations during non-working hours effectively reduces the production of cosmogenic radionuclides by about a factor of 3.
The cosmogenic activation in germanium during transportation is well mitigated by the transportation shield, so their specific activities grow slowly and even show a downward trend due to the natural decay. 
At the final stage, HPGe detectors are transported to CJPL and all cosmogenic radionuclides stop accumulating underground. The specific activities of $^{68}$Ge and $^{60}$Co are 17.53 and 0.10 $\mu$Bq/kg in this moment.

\section{Cosmogenic background assessment in HPGe detectors}\label{sec.III}

\begin{figure}[htbp]
\centering
\includegraphics[width=.8\hsize]{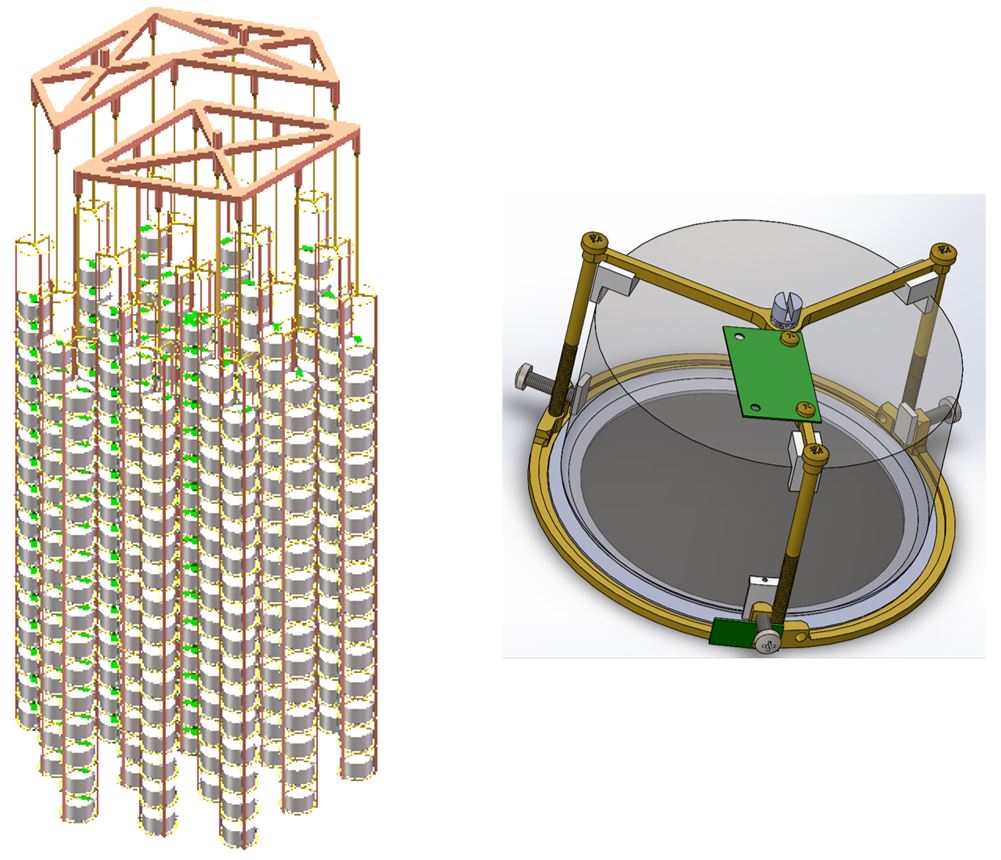}
\caption{The schematic diagram of the CDEX-300$\upsilon$ with 19 strings of 16 detectors (left) and HPGe detector unit (right) of the CDEX-300$\upsilon$ in SAGE.}
\label{fig4}
\end{figure}

As shown in figure~\ref{fig4}, CDEX plans to establish a hexagonal detector array containing 300 kg germanium at the second phase of CJPL for CDEX-300$\upsilon$ experiment. The HPGe detector units, enriched in $\ge86\%$ $^{76}$Ge, will be deployed in 19 strings. Materials used in detector fabrication will be measured and selected to meet the background goal. The detector array will be operated in a 1725 m$^{3}$ cryostat and surrounded by over 6 m thick liquid nitrogen, which can provide an effective shield against environmental radioactivity. Under these background control measures, the background from primordial radionuclides will be controlled at a very low level.

\begin{figure}[htbp]
\centering
\includegraphics[width=.8\hsize]{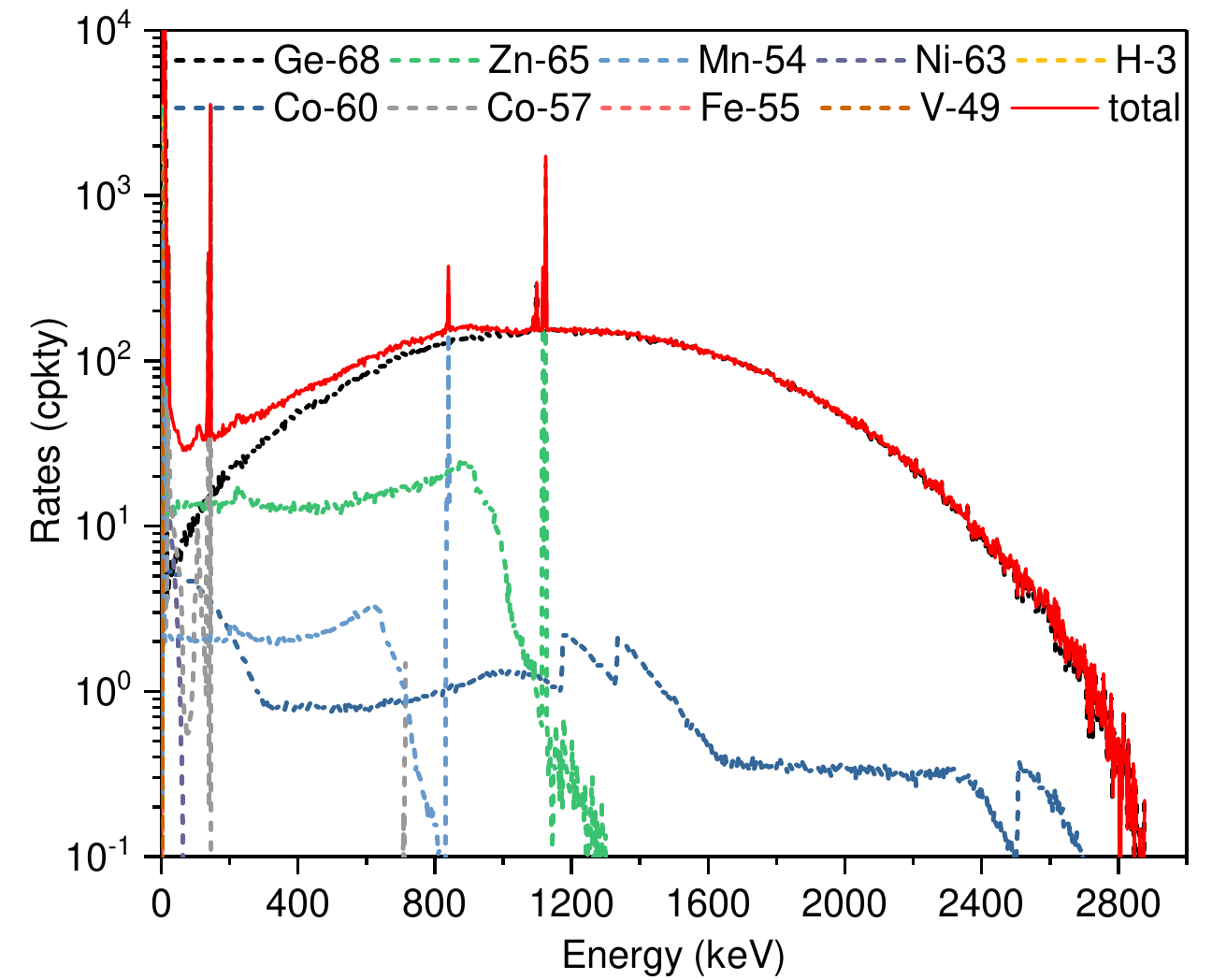}
\caption{The simulated background spectra from cosmogenic radionuclides in germanium for the CDEX-300$\upsilon$ with a cooling time of 1 year. The energy resolution is from the germanium detector used in CDEX-1B \cite{b10}.}
\label{fig5}
\end{figure}

CJPL has a rock overburden of about 2400 meters, which almost completely suppresses the cosmogenic activation of germanium \cite{b27}. HPGe detectors will be stored in CJPL for about 1 year, so-called cooling time, to prepare for the experiments and reduce the background through the natural decay of cosmogenic radionuclides. With the initial specific activities obtained in section~\ref{sec.II}, the background contribution of cosmogenic radionuclides in germanium was estimated with a Geant4-based Monte Carlo framework, called Simulation and Analysis of Germanium Experiment (SAGE)\cite{b28}. In the simulation, cosmogenic radionuclides were evenly distributed in all germanium crystals. The energy spectra of the cosmogenic radionuclides in germanium were simulated as shown in figure~\ref{fig5}.

The CDEX-300$\upsilon$ experiment aims to achieve a discovery potential that reaches the inverted hierarchy mass region with 1 ton·yr exposure ($\thicksim$5 yr running). The half-life sensitivity of the experiment is mainly determined by exposure time and the background rate around 2039 keV ($Q_{\beta \beta}$ of $0\upsilon\beta\beta$ decay of $^{76}$Ge). The total background rate from cosmogenic radionuclides in germanium is simulated to be 49.59 counts per keV per ton per year (cpkty) in the energy range of 2020$\thicksim$2060 keV after 1 yr cooling at CJPL, 99\% of which is contributed by $^{68}$Ge. Cosmogenic activation induced by cosmic-ray neutrons contributes 81\% of $^{68}$Ge production. 
Additional neutron shielding is beneficial to mitigate the background contribution from $^{68}$Ge. According to simulations, 40 cm thick polyethylene cover can further reduce the production rates of $^{68}$Ge by about a factor of 2. 
Besides, appropriate cooling time can significantly mitigate the background contribution from $^{68}$Ge. When the cooling time is extended to 2 years, the specific activity of $^{68}$Ge will decrease by 60.69\% and the total background rate from cosmogenic radionuclides in germanium will decrease by 60.36\%.

\section{Summary} \label{sec.IV}
The background induced by cosmogenic activation of germanium above ground was simulated according to the scheduled fabrication and transportation processes for the $^{76}$Ge enriched HPGe detectors of the next generation CDEX experiment. The production rates of cosmogenic radionuclides in germanium were simulated with Geant4 and the CRY-1.7 library. The production rates of two typical cosmogenic radionuclides, $^{68}$Ge and $^{60}$Co, during detector fabrication were 13.45 and 1.23 kg$^{-1}$d$^{-1}$, respectively. Based on the simulated results, we assessed the background contribution from cosmogenic radionuclides for CDEX-300$\upsilon$ experiment with the SAGE code. The total background rate from cosmogenic radionuclides in germanium was simulated to be 49.59 cpkty in the energy range of 2020$\thicksim$2060 keV for $0\upsilon\beta\beta$ search after 1 yr cooling at CJPL, dominated by the decay of $^{68}$Ge.
Additional neutron shielding and appropriate cooling time can further mitigate the background contribution from $^{68}$Ge.

\acknowledgments

This work was supported by the National Key Research and Development Program of China (Grants No. 2022YFA1604701) and the National Natural Science Foundation of China (Grants No. 12175112).

% We suggest to always provide author, title and journal data:
% in short all the informations that clearly identify a document.

\bibliographystyle{JHEP}
\bibliography{JINST_Manuscript_nqy}

\end{document}